\documentclass[12pt]{article}
\usepackage{graphicx}
\usepackage{epsfig}
\usepackage{amsmath}
\usepackage{amsfonts}
\usepackage{amssymb}
\begin{document}
\begin{titlepage}
\hfill
\begin{tabular}{l}
\end{tabular}
\vspace{1in}
\begin{center}
{\LARGE Large-$x$ Parton Distributions}\\
\vspace{1in}
{\large S.~Kuhlmann$^a$, J.~Huston$^d$, J.~Morfin$^b$,
F. Olness$^f$, \ J.~Pumplin$^d$, J.~F.~Owens$^c$,
W.~K.~Tung$^d$ and J.~J.~Whitmore$^e$ \\}
\vspace{1.25cm}
{\large
$^a$Argonne National Laboratory, $^b$Fermi National Accelerator Laboratory, \\
$^c$Florida State University, $^d$Michigan State University, \\
$^e$Pennsylvania State University, $^f$Southern Methodist University}
\vspace{1in}
\end{center}
\begin{abstract}
Reliable knowledge of parton distributions at large $x$ is crucial for
many searches for new physics signals in the next generation of collider 
experiments.
Although these are generally well determined in the small and medium $x$ range,
it has been shown that their uncertainty grows rapidly for
$x > 0.1$. We examine the status of  the gluon and quark
distributions in light of  new questions that have been raised in the past
two years about ``large-$x$'' parton distributions, as well as recent 
measurements
which have improved the parton uncertainties.  We perform several specific
global fits to answer the outstanding questions, and to investigate the
potential of future experiments, particularly at HERA, which may provide
critical constraints on parton distributions at large $x$.
\end{abstract}
\vfill
\end{titlepage}

\section{Introduction\bigskip}

Four years ago the CDF collaboration reported \cite{cdfjet} an excess of jet
events at large transverse energy over perturbative Quantum Chromodynamics
(QCD) calculations.  A possible explanation for this effect was a larger than
expected gluon distribution at large $x$ \cite{cteqjet}. Three years ago the
deep-inelastic scattering (DIS) experiments at HERA reported a low statistics
excess of events at large Q$^{2}$ \cite{hera}.  This led to speculation
that part of this excess could be attributed to a lack of knowledge of the
quark distributions at large $x$ \cite{highxq}, and could possibly be
related to the jet events which are produced by a combination of quark and
gluon scattering.  Both excesses produced a large number of papers about
the possible implications for physics beyond the Standard Model,  emphasizing
the need for much better knowledge of parton distributions at large $x$.

In the past few years there has been considerable progress towards
understanding some of the uncertainties in the individual measurements that
contribute to our knowledge of large-$x$ parton distributions (PDFs),  but in
some cases this has led to an \emph{increase} in the uncertainty of the
large-$x$ PDFs,  rather than a reduction.   This letter will review the
recent analyses and point towards future measurements which may help clarify
the situation.   First we must better define ``large $x$''.    For the
gluon distribution there is little confusion:  gluon distributions for
$x>0.1$ at all Q$^{2}$ become increasingly uncertain as shown in ref.
\cite{scan},  and further described below.   The quark distributions are
more complicated due to a strong Q$^{2}$ and flavor dependence.   The
incoherent sum of quark distributions at moderate Q$^{2}$ ($\approx$25-1000
GeV$^{2})$ is known to be well understood up to $x\approx0.7,$  but the
earlier speculations \cite{highxq} concerned large $x$ ($x>0.5$) and large
Q$^{2}$ (%
$>$%
10000 GeV$^{2}$).  In addition,  when one examines the individual flavors of
quark distributions, the uncertainties grow significantly for $x>0.3$.   All
of these issues will be discussed in detail,  beginning with the gluon 
distribution.

\section{\bigskip Gluon Distribution}

In the past few years there has been very little progress in reducing the
uncertainty in the gluon distribution at large $x$,  and new questions have
arisen which perhaps confuse the situation even more.  The most recent
analyses of the gluon distributions from the CTEQ5 \cite{cteq5} and MRST
\cite{mrst} collaborations have reinforced the two conclusions of the earlier
CTEQ4 gluon parameter scan \cite{scan},  1) that the gluon uncertainties for
$x<0.1$ are reasonably small and of order 10\%,  and 2) for $x>0.1$ the
uncertainties grow significantly.   This is illustrated in fig.1,  which
shows the ratio of gluon distributions at Q=100 GeV to that of the CTEQ4M
gluon distribution \cite{cteq4m}.   The solid lines include both of the
CTEQ5 gluon distributions (CTEQ5M and CTEQ5HJ),  as well as the gluon
distributions from the parameter scan mentioned above.  The dashed lines are
the three gluon distributions from the MRST analysis.  One notices that the
``bands'' from CTEQ and MRST are quite consistent at low $x$,  but barely
overlap at large $x$.   

What features of the CTEQ and MRST analyses cause the gluon discrepancy at
large $x$?   This is due to the choices of different data sets used in each
analysis,  in particular the emphasis on Tevatron jet data by CTEQ and direct
photon data by MRST.   In addition,  the specific treatment of the direct
photon data with respect to the issue of k$_{t}$ smearing \cite{joeykt} plays
a significant role.   The CTEQ and MRST groups agree that the direct photon
theory needs some kind of correction for k$_{t}$ smearing,  but without a
full theoretical framework the procedures for deriving such corrections are
somewhat arbitrary, and are significantly different between ref. \cite{joeykt}
and the MRST analysis.   Therefore there are three scenarios that result in
very different gluon distributions: 1) Emphasis on the CDF collider jet data,
 which leads to the CTEQ5HJ set of parton distributions.   These can be
considered an upper bound to the gluon distribution at large $x$.   2)
Emphasis on direct photon data and the k$_{t}$ correction procedure of MRST.
 These can be considered a lower bound at large $x$.   3) Emphasis on
direct photon data but using the k$_{t}$ procedures of ref. \cite{joeykt}.
 This yields gluon distributions which are consistent with those such as
CTEQ5M which tend to lie halfway between the two bounds.   In addition,
there are other complications in using both the direct photon and jet data
sets.   As pointed out in refs. \cite{mrst}\cite{joeykt},  it is difficult
to reconcile the k$_{t}$ values needed by the WA70 and E706 experiments.
  The jet data, on the other hand,  only used at transverse energies larger
than 40 GeV,  should not be significantly affected by k$_{t}$.   These
measurements and comparisons to theory have their own set of concerns.
 Included in these is the definition of the jet,  which can never be exactly
the same in the data and in a next-to-leading-order QCD calculation.   The
precise statistical procedure for the jet data, which are dominated by highly
correlated systematic uncertainties,  is another area of concern
\cite{lyons}.  

Clearly much more work is needed on the gluon distribution at large $x$:
 what can be done to improve the constraints?     Obviously the best
scenario is a complete understanding of soft gluon effects in the direct
photon data sets,  in which case the E706 data are sufficiently precise to
severely constrain the gluon distribution.    This may take many years,
however,  and the discrepancies between data sets may never be understood.
  In addition,  more work is needed on the Tevatron jet data and their
interpretation,  especially the recent differential dijet measurements from
both CDF and D0.   These data are also sensitive to changes in the quark
distributions, discussed in the next section.   Finally,  another
possibility for a future measurement is to use the Drell-Yan process to
measure the large-$x$ gluon distribution at the Fermilab Main Injector, as
discussed in ref. \cite{klasen}.   Acquiring the needed data set for this
measurement is more speculative, but appears to be worth a serious study of
the potential of such an experiment.

\section{\bigskip General Quark Distribution Issues}

There are four main ways that large-$x$ quark distributions may be modified in
a significant way (enough to affect Tevatron and/or HERA processes),  while
maintaining agreement with fixed target data sets:    1) a modification of
the u quark near $x\approx1$,  2) a non-perturbative ``intrinsic charm'' type
of component that is presently assumed to be zero,  3) a modification of the
d quark at large $x$, and 4) higher twist contributions that are missing from
the conventional fits to the low Q$^{2}$ fixed target data.   The first
three were discussed in ref. \cite{highxq},   where an example toy model of
a modified u quark distribution was presented.   Much more is now known
about the constraints on such models,  as will be discussed next.   The d
quark issues will be discussed in detail in the next section.

The first ``new'' constraint is the reanalysis of large-$x$ SLAC electron
scattering data off hydrogen targets,  discussed in ref. \cite{yang}.
  This data set is in the resonance region and one must assume that the
Bloom-Gilman duality hypothesis \cite{bloom} can be applied.   In addition,
 these data require target mass corrections that are enormous, up to a factor
of 50 near $x\approx1$.  The target mass corrections are mostly derived by
using the Nachtmann scaling variable instead of $x$ \cite{yang},  and are a
fairly straightforward kinematic shift due to the mass of the proton.   The
duality hypothesis and target mass corrections are probably accurate enough to
constrain modification \#1 above,  but one would like another
measurement/process to confirm these assumptions.   In addition,  these
data are below the charm threshold and therefore say nothing about intrinsic
charm models.    A recent neutrino scattering measurement from
CCFR \cite{ccfrhix} appears to provide some confirmation of the electron
analysis.   These data are also at higher Q$^{2}$ and therefore above the
charm threshold.  A concern with this measurement is the nuclear effects from
the iron target.   But with relatively simple Fermi motion corrections,
 the data are within a factor of 2 of predictions using conventional parton
distributions of the nucleon such as CTEQ4M.    The comparison can be
further improved with more sophisticated treatments of the nuclear effects.
 The combination of the neutrino and electron data analyses makes it unlikely
that either of the first two modifications of the quark distributions are
large enough to affect collider measurements.  

Phenomenological fits for higher twist effects are described in refs.
\cite{mrstwist}\cite{bodektwist},  while one theoretical model for the
parton-parton correlations involved is discussed in ref. \cite{qiu}.   Both
the model and the fits show $\approx$3\% changes in the valence quark
distributions in the $0.1<x<0.5$ range,  growing to 5-10\% changes at
$x\approx0.8$.   The changes described in these papers should also be
considered as part of the uncertainty in the parton distributions.

\section{\bigskip d/u Ratio}

 The ratio of the density of down quarks to that of up quarks in the proton
has changed in the most recent CTEQ and MRST analyses due to the new W
lepton-asymmetry data from CDF \cite{wasym}, as well as the NMC ratio
measurement of deuterium/hydrogen scattering \cite{nmc}.   For many years
the basic assumptions about the parameterization of this ratio and the use of
the DIS data have been relatively unchallenged,  but this has changed.
  The two main reasons to question these assumptions are: 1) the behavior of
the d/u ratio as $x\rightarrow1$,  and 2) possible nuclear binding effects
in the deuteron.   We will now review some of the history of these two issues.

The extrapolation of the d/u ratio was discussed in non-perturbative
QCD-motivated models in the 1970's such as ref. \cite{farrar}.  These models
predicted that the ratio should approach 0.2 as $x\rightarrow1$.   Other
models predicted the ratio should go to zero; but since neither is
convincing the asymptotic value of this ratio has been set arbitrarily by the
choice of parameterizations of the CTEQ and MRS groups.   The choices that
were made drive the ratio to zero as $x\rightarrow1$. Some papers in recent
years,  such as ref. \cite{thomas},  have called for a special set of parton
distributions that force the ratio to 0.2 as an alternative to the standard
sets.   Such a fit has now been performed and will be discussed below.

More than five years ago the SLAC experiment E139 published a series of
measurements \cite{slac} with different targets.  One of the goals of the
measurement was to see if nuclear binding effects were present in deuterium.
 This was accomplished by a global fit to all the target data,  within the
context of a non-perturbative nuclear density model \cite{frankfurt}.   The
conclusion was that the binding effects clearly seen in heavier nuclei are
also present in deuterium at the few percent level.  This result is not
surprising, and is perhaps even expected,  but it is not conclusive for two
reasons:  1) it depends critically on an unproven nuclear physics model with
many parameters that had to be obtained from fits to the data,  and 2) the
deuteron is a very special nucleus with binding energies much smaller than
the rest, so that a large extrapolation from the heavier nuclei is needed.
 Ref. \cite{duwrong} argues that a proper extrapolation predicts no binding
effects in the deuteron.   With the caveats just mentioned we consider the
corrections to have a large and unquantified uncertainty.

The effects in the previous two paragraphs were ignored until the analysis of
Yang and Bodek \cite{yang} two years ago.  They took the latest W
lepton-asymmetry and NMC ratio data and proposed a modification of the d/u
ratio that included the nuclear binding effects and forced the ratio to 0.2 as
$x\rightarrow1$.    This proposal has fueled considerable interest in these
issues.   However, the paper implied that the W lepton-asymmetry and NMC
ratio data  could be fit  \textit{only }with the nuclear binding corrections
\textit{and }with d/u $\rightarrow0.2$ as $x\rightarrow1$,   which we will
show is not the case.  In fact, both data sets can be fit quite well without
\textit{either} modification\textit{.}

To illustrate the different possibilities, a new series of fits was performed
within the context of the CTEQ5 global analysis \cite{cteq5}.   The nuclear
binding corrections were included as well as fits with a modified behavior of
d/u as $x\rightarrow1$.  We find we can get a good fit to all the data with
neither correction, or with the nuclear binding corrections added but with any
d/u behavior as $x\rightarrow1$.   Figures 2 and 3 show examples of this.
 Figure 2 shows the NMC ratio data with and without the deuteron correction.
 The lower (solid) curve is CTEQ5M,  while the upper (dashed) curve is a new
fit to the corrected data, again with the standard CTEQ5 parameterization
which forces d/u to zero as $x\rightarrow1$.   Both are good fits to the NMC
data, as well is a new third option (not shown since it lies precisely on the
dashed curve) which includes both the nuclear corrections and the changed d/u
parameterization.  

 Figure 3 shows the d/u ratio resulting from these three fits at Q=80 GeV
(there is very little evolution dependence in this ratio).  All three are
viable candidates for the d/u ratio,  and the upper and lower ones could
quite reasonably be considered upper and lower bounds.  Figure 3 also
includes vertical lines to distinguish the three regions of $x$ involved in
this study,  and to help explain why the different effects can be treated as
independent.   For $x<0.3$ the W lepton-asymmetry data and the NMC ratio
data are both very precise \textit{and }the nuclear corrections to the NMC
data are insignificant.  The two measurements agree so the d/u ratio is very
well constrained in this region.  Unfortunately the present W asymmetry data
end near $x=0.3$,   precisely where the nuclear corrections to the NMC data
become significant.  Therefore with any reasonably flexible parameterization
one can get a spread of d/u ratios for $0.3<x<0.7$ (the middle region of the
plot) simply by changing the nuclear correction,  and still fitting the W
asymmetry and NMC data.   Finally for the largest $x$ values, we note that
the NMC data end near $x=0.7;$  therefore many different extrapolations to
$x\rightarrow1$ are possible, with or without nuclear corrections.  Clearly
the issues for the three different regions are quite independent.  

It is worth noting that if d/u$\rightarrow0.2$ as $x\rightarrow1$,  then
there \textit{must }be some nuclear corrections in order to fit the NMC data.
 The previous discussion shows that the converse is not necessarily true.
  However if appreciable binding effects are present in the deuteron,  then
it is perhaps more natural for d/u to go to a constant than to zero,  which
would require a fairly sharp downturn near $x=1$.   Assuming that d/u does
not suddenly increase as $x\rightarrow1$,  this constant is unlikely to be
larger than 0.22,  since that is where the last NMC data point lies.   But
any constant between 0.05 and 0.2 would be a reasonable extrapolation and is
not constrained by present data.  The only bias is the theoretical one
mentioned earlier for 0.2,  which we do not find persuasive. 

One possible way to constrain the d quark is from measurements of $\pi^{+}%
/\pi^{-}$ production in DIS interactions, as described in ref. \cite{piondu}.
  But certainly the best way to constrain the d quark in the future, in
terms of both experimental and theoretical uncertainties, is with high
luminosity HERA measurements of positron-induced charged current
interactions.   The upper two plots in fig. 4 show the most recent H1
\cite{h1cc} and ZEUS \cite{zeuscc} charged current measurements.   For the
H1 data the cuts are Q$^{2}$%
$>$%
1000 GeV$^{2}$and $y$%
$<$%
0.9, while for ZEUS the cut is Q$^{2}>200$ GeV$^{2}.$  They are compared to a
NLO QCD calculation using the standard CTEQ5D (DIS scheme) set of parton
distributions,  which are fit without the binding corrections to the NMC
data.   The lower two figures show the ratios (solid curves) with respect to
the theory using CTEQ5D.  This provides a good description of the data,
 although there is a hint of a low statistics excess in the ZEUS data.
   The dashed curves in the ratio plots are a second DIS scheme fit, which
we label CTEQ5DU, including binding corrections but with the CTEQ5
parameterization (d/u $\rightarrow0$) corresponding to the dashed curve (in
the $\overline{MS\text{ }}$ scheme) in fig. 3.   Since the data are below
$x<0.7$ the fits with d/u$\rightarrow0.2$ give the same result as CTEQ5DU in
this plot.   Parton distributions similar to the dashed and solid curves
were used to estimate the required luminosity to distinguish them.  The
result is that 500 pb$^{-1}$ of delivered positron luminosity (250 pb$^{-1}$
in each of the two experiments) \cite{zeus} is needed to achieve a 2 standard
deviation separation.   This is clearly a large data set but not impossible
with the forthcoming HERA upgrade.  We think it is vital that the HERA
program continue until this issue is settled.

\section{\bigskip Conclusions and Acknowledgements}

In the past four years new information has become available concerning
large-$x$ parton distributions and their uncertainties.  The issues have
become more important with the realization that these uncertainties could be
hampering searches for physics beyond the Standard Model.   The different
analyses reviewed in this letter have clarified some of the issues but have
also raised new questions to be addressed.  We have outlined a program of
measurements at the Tevatron and HERA colliders,  as well as important
theoretical work, that is needed to improve the uncertainties in large-$x$
parton distributions before the LHC era begins.

We wish to thank M. Kuze, K. Nagano, and A. van Sighem for many useful
discussions and for the HERA luminosity estimate needed to determine the d/u 
ratio.

\newpage

\begin{figure}[ptbh]
\epsfig{file=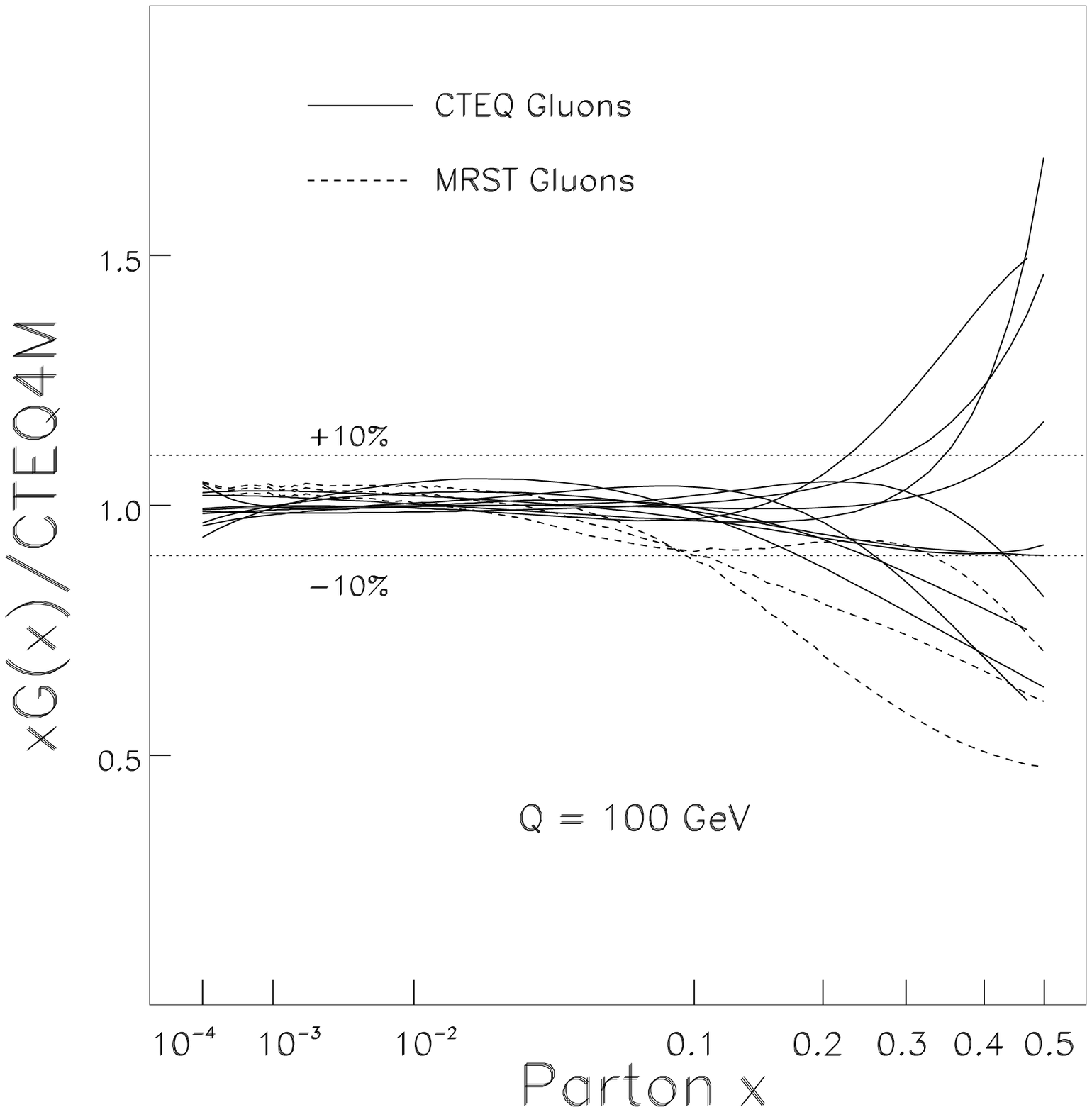,width=\linewidth}\caption{Ratios of gluon
distributions are shown, compared to the CTEQ4M gluon distribution. The solid
lines are distributions from the CTEQ4 and CTEQ5 analyses, and the dashed are
from the MRST analysis (see text).}%
\end{figure}

\begin{figure}[ptbh]
\epsfig{file=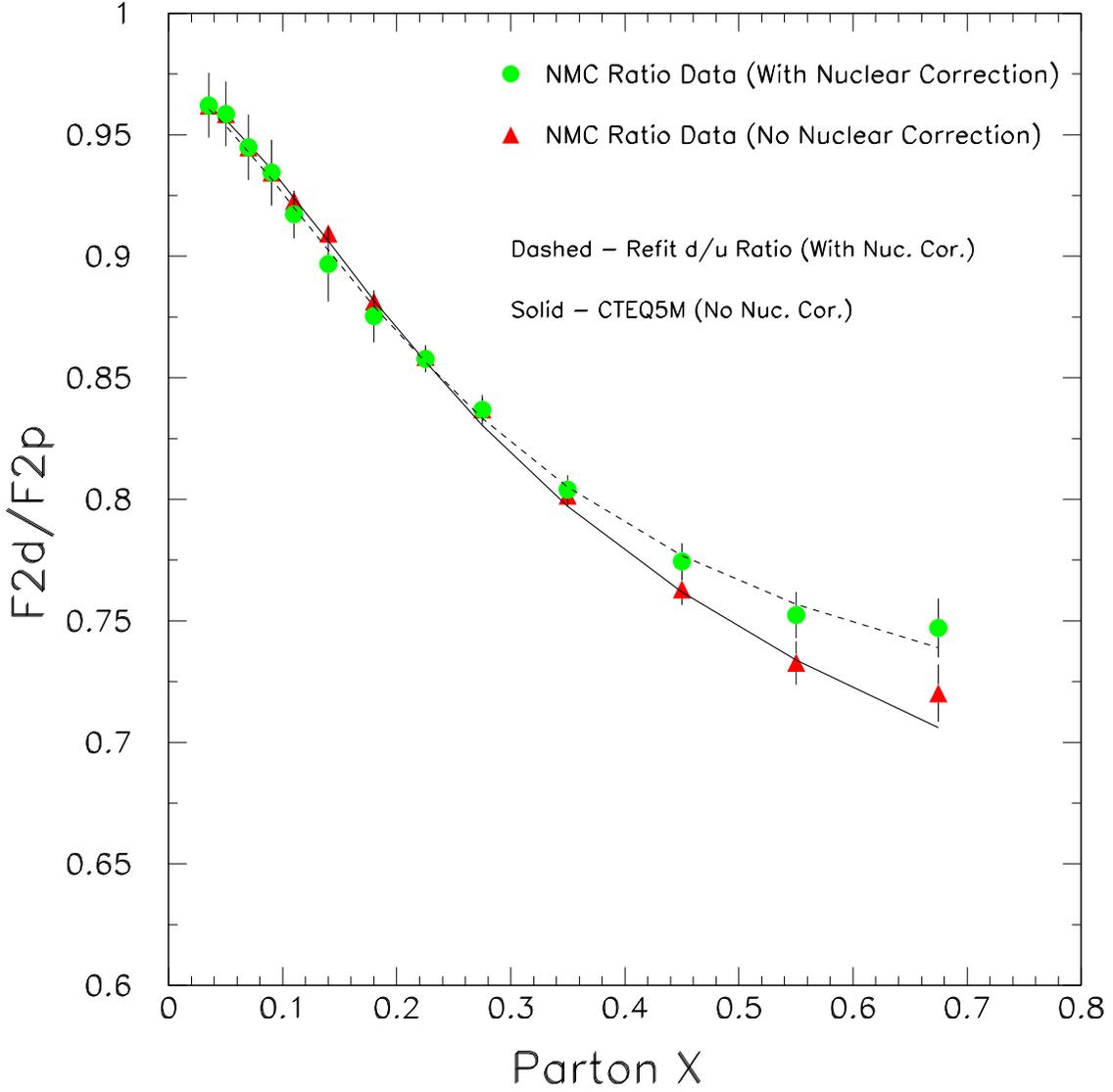,width=\linewidth}\caption{The measured ratio of
muon scattering off deuterium and hydrogen targets, from the NMC experiment,
is shown with and without the nuclear corrections described in the text. The
solid line is CTEQ5M which was fit to the data with no nuclear corrections,
while the dashed line is a fit to the data with the nuclear corrections,
altering the d/u ratio (see text).}%
\end{figure}

\begin{figure}[ptbh]
\epsfig{file=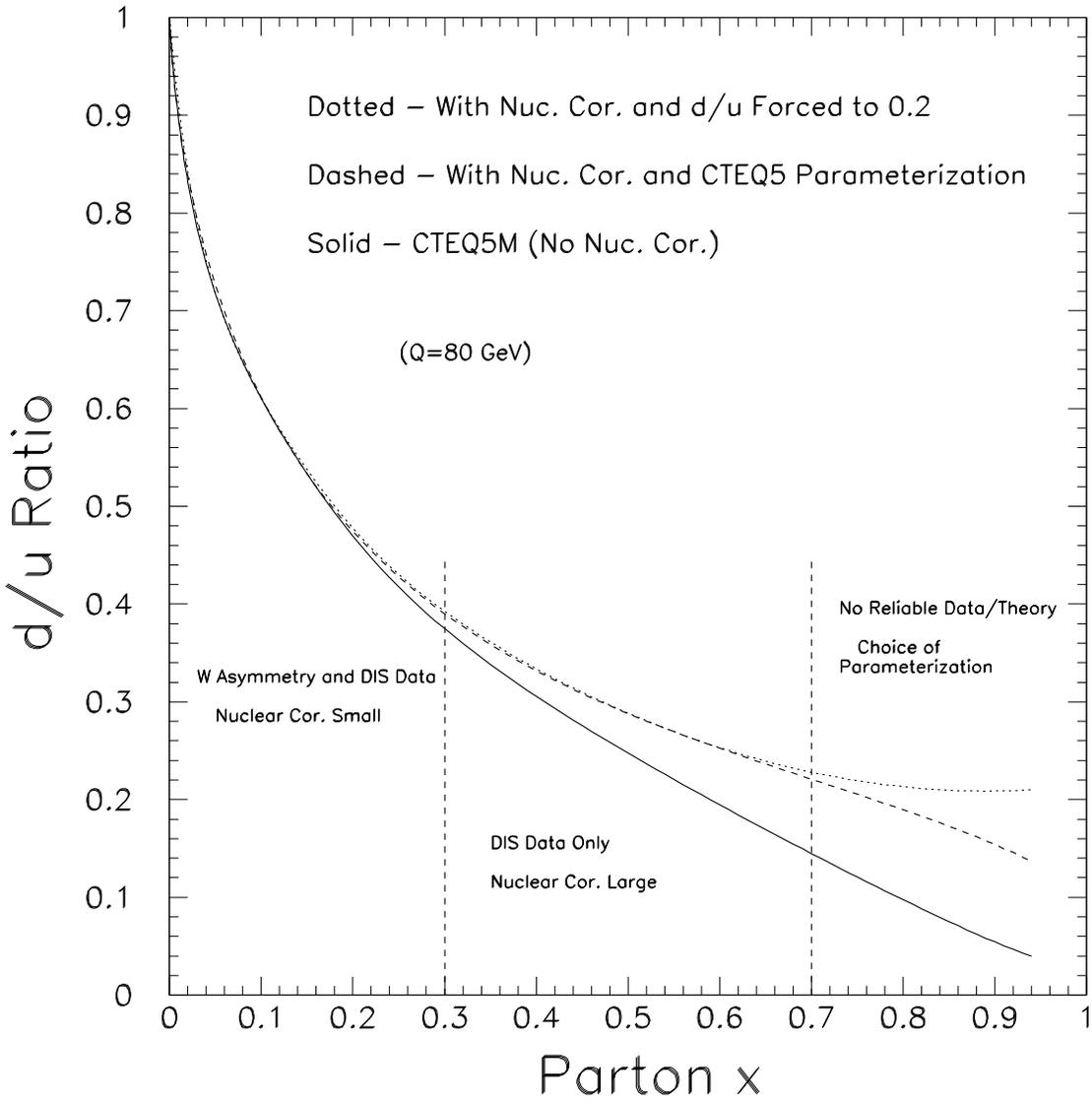,width=\linewidth}\caption{The d/u ratio is shown for
the three global fits described in the text. Also shown are the three
different regions of $x$ and the relevant measurements in each region.}%
\end{figure}

\begin{figure}[ptbh]
\epsfig{file=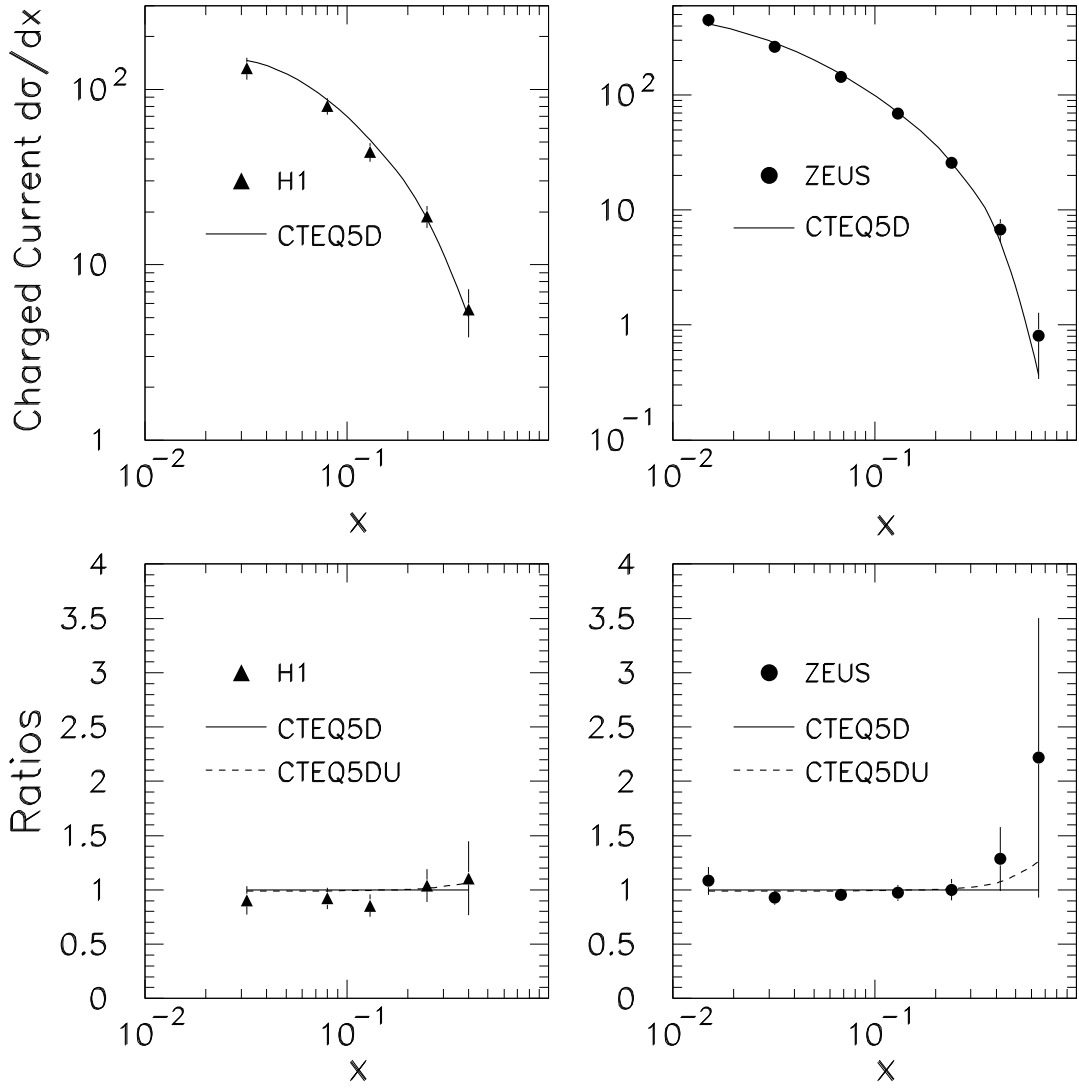,width=\linewidth}\caption{Positron-induced charged
current data from H1 and ZEUS are shown, along with NLO QCD calculations using
parton distributions fit with and without nuclear corrections to the fixed
target data (see text).}%
\end{figure}
\end{document}